\documentclass[12pt]{iopart}
\newcommand{\gguide}{{\it Preparing graphics for IOP journals}}
\usepackage{latexsym}
\usepackage{graphics}
\usepackage{xcolor}

\begin{document}

\title[Characteristic functions under series and parallel connection of quantum graphs]
{Characteristic functions under series and parallel connection of quantum graphs}

\author{ V Pivovarchik}

\address{South Ukrainian National Pedagogical University,
Staroportofrankovskaya str., 26, Odessa 65020, Ukraine}
\ead{v.pivovarchik@paco.net}
\begin{abstract}
For a graph consisting of parallel connected subgraphs we express the characteristic function of the  boundary value
problem with generalized Neumann conditions at both joining points via characteristic functions of different boundary problems on the subgraphs.

\end{abstract}

\vspace{2pc} \noindent{\it Keywords}: Dirichlet boundary
condition, Neumann boundary condition, Kirchhoff condition, continuity condition, subgraph, connectivity, boundary value problem,
spectral parameter.

\submitto{Journal of Physics: A}
\maketitle

\section{Introduction}
\setcounter{equation}{0} \hskip0.25in

We consider boundary value problems generated by the Sturm-Liouville equation on a connected compact metric graph with continuity and Kirchhoff's conditions at interior vertices and Robin or Dirichlet conditions at pendant vertices  (i.e. vertices of degree 1).  The potentials in the Sturm-Liouville equations and constants in the conditions are assumed to be real, and such that the corresponding operator is self-adjoint. 

To describe the spectrum of a boundary value problems on metric graphs it is possible to use characteristic functions or spectral determinants. 

It was proved in \cite{LP} that the characteristic function $\Phi_N(\lambda)$ of the boundary value problem on a graph $G$ which consists of two subgraphs $G_1$ and $G_2$ with the generalized Neumann (i.e. continuity + Kirchhoff's) boundary condition at the only cut-vertex{ (see \cite{Tut}, p.54 for a definition)}{ \bf v} of $G$ satisfies  
\begin{equation}
\label{1.1}
\Phi_{N}(\lambda)=\Phi_N^1(\lambda)\Phi^2_D(\lambda)+\Phi_D^1(\lambda)\Phi_N^2(\lambda),
\end{equation}
where $\Phi_N^j(\lambda)$ ($j=1,2$) are the characteristic functions of boundary value problems on subgtaphs $G_1$ and $G_2$ with  Neumann conditions at {\bf v}, and  $\Phi_D^j(\lambda)$ are the characteristic functions of boundary value problems on $G_1$ and $G_2$ with Dirichlet condition at {\bf v}.

Formula (\ref{1.1}) was proved in \cite{LP} for trees but the proof is quite the same for any separable graph. Earlier this formula was obtained for the so-called
spectral determinants \cite{T}. In  case of Neumann conditions at pendant vertices and under the condition of no loops and multiple edges the spectral determinant is nothing but our characteristic function. Also it should be mentioned that in \cite{T} a definition of 
spectral determinant introduced in \cite{D} was used which is not quite accurate. The solutions $\psi_{\alpha\beta}$ and $\psi_{\beta\alpha}$ defined by (1)--(3) in \cite{D}  not always  exist, e.g. in case of $\gamma=\alpha=0$, $\beta=\pi$ (in terms of \cite{D}) these solutions are absent.  Therefore  we use the characteristic functions language in  the present paper to be mathematically regorous. Formula (\ref{1.1}) for the graph $P_2$ (a path with two edges) has been used for solving the so-called inverse three spectra problem \cite{P} and the Hochstadt-Lieberman problem \cite{MP} and for describing  spectral problems generated by equations of Stieltjes strings  on trees in \cite{LPW},  \cite{PRT}, \cite{PT}.

A graph of connectivity 1 (see \cite{De}, p.75 or  \cite{Tut}, p.72 for a definition), i.e. a separable graph can be considered as a series connection of its subgraphs. In the present  paper we
consider a parallel connection of subgraphs into a graph of connectivity 2 aiming to deduce an analogue of (\ref{1.1}) for parallel connection of subgraphs.

\section{Characterictic functions
}
\setcounter{equation}{0} \hskip0.25in

Characteristic functions of boundary value problems on graphs are natural generalizations of characteristic functions of boundary value problems on an interval. For an interval we use the following four characteristic functions. If $s(\lambda,x)$ is the solution of the Sturm-Liouville equation with a real valued potential $q(x)\in W_2^2(0,l)$:
\begin{equation}
\label{2.1}
-y^{\prime\prime}+q(x)y=\lambda^2y, \ \ \ x\in(0,l), 
\end{equation}
which satisfies the conditions $s(\lambda,0)=s^{\prime}(\lambda,0)-1=0$ and $c(\lambda,x)$ is the solution which satisfies $c(\lambda,0)-1=c^{\prime}(\lambda,0)=0$, then  we call $\Phi_{DD}(\lambda)\mathop{=}\limits^{def}s(\lambda,l)$ the {\it Dirichlet-Dirichlet} characteristic function because the set zeros of $\Phi_{DD}(\lambda)$ coincides with the spectrum of the 
problem generated by (\ref{2.1}) and the boundary conditions
$$
y(0)=y(l)=0.
$$
We call $\Phi_{DN}(\lambda)\mathop{=}\limits^{def}s^{\prime}(\lambda,l)$ the {\it Dirichlet-Neumann} characteristic function because the set of zeros of $\Phi_{DN}(\lambda)$ coincides with the spectrum of the 
problem generated by (\ref{2.1}) and the boundary conditions
$$
y(0)=y^{\prime}(l)=0,
$$
we  call $\Phi_{ND}(\lambda)\mathop{=}\limits^{def}c(\lambda,l)$ the {\it Neumann-Dirichlet} characteristic function because the set of zeros of $\Phi_{ND}(\lambda)$ coincides with the spectrum of the 
problem generated by (\ref{2.1}) and the boundary conditions
$$
y^{\prime}(0)=y(l)=0,
$$
we  call $\Phi_{NN}(\lambda)\mathop{=}\limits^{def}c^{\prime}(\lambda,l)$ the {\it Neumann-Neumann} characteristic function because the set of zeros of $\Phi_{NN}(\lambda)$ coincides with the spectrum of the 
problem generated by (\ref{2.1}) and the boundary conditions
$$
y^{\prime}(0)=y^{\prime}(l)=0.
$$

Now let us consider boundary value problems on a compact connected graph.
Let $G$  be a metric graph with $g$ edges. We denote by $v_j$ the vertices of $G$, by $d(v_j )$ their degrees,
by $e_j$ the edges of $G$ and by $l_j$ their lengths. An arbitrary vertex v is chosen as the root. Since as it will be clear below our results do not depend on the orientation of the edges, we fix an arbitrary orientation but for the sake of convinience we assume that  the root {\bf v} has only outgoing  incident edges. 
Local
coordinates for edges identify the edge $e_j$ with the interval $[0, l_j ]$ so that the local coordinate
increases in the direction of the edge. This means that the root {\bf v}  has the
local coordinate $0$ on each incident edge.  All the other interior vertices $v$ may have  outgoing edges, with local coordinates $0$, and incoming edges $e_j$ with local coordinates $l_j$ . Functions $ y_j$ on the edges are subject
to the scalar Sturm- Liouville equations:
\begin{equation}
\label{2.2}
−y^{\prime\prime}_
j + q_j (x_j)y_j = \lambda^2y_j , 
\end{equation}
where $q_j$ is a real-valued function which belongs to $L_2[0, l_j ]$. For an edge $e_j$ incident to a
pendant vertex which is not the root we impose self-adjoint boundary conditions
\begin{equation}
\label{2.3}
y^{\prime}_
j (0) + \beta_j y_j (0) = 0, 
\end{equation}
or
\begin{equation}
\label{2.4}
y^{\prime}_j (l_j) + \beta_j y_j (l_j) = 0, 
\end{equation}
where $\beta_j\in R\cup \{\infty\}$. The case $\beta_j = \infty$ corresponds to Dirichlet boundary condition
$y_j (0) = 0$ or $y_j(l_j)=0$.

 For each  interior vertex which is not the root  with
incoming edges $e_j$ and outgoing edges $e_k$ the continuity conditions are
\begin{equation}
\label{2.5}
y_j (l_j ) = y_k(0), 
\end{equation}
for all $j$ and $k$ and  Kirchhoff's condition is
\begin{equation}
\label{2.6}
\sum\limits_ky_k^{\prime}
(0) =\sum\limits_jy_j^{\prime}(l_j).
\end{equation}

At the root ${\bf v}$, we impose the continuity conditions
\begin{equation}
\label{2.7}
y_j (0 ) = y_k(0)
\end{equation}
for all  edges $e_j$ and $e_k$ outgoing from ${\bf v}$  and Kirchhoff's condition
\begin{equation}
\label{2.8}
\sum\limits_ky_k^{\prime}
(0) =0
\end{equation}
where the sum in the left-hand side is taken over all edges  outgoing from $\bf{v}$. 

We will call  the pairs of conditions (\ref{2.5})-(\ref{2.6}) and    (\ref{2.7})-(\ref{2.8})      {\it generalized Neumann conditions} for an interior vertex. It is clear that being imposed at a pendant vertex these conditions are reduced to the usual Neumann condition (condition (\ref{2.3})  or (\ref{2.4}) with $\beta_j=0)$.

Let us denote by $s_j (\lambda, x_j)$ the solution of the Sturm-Liouville equation (\ref{2.2}) on an
edge $e_j$ which satisfies the conditions $s_j (\lambda, 0) = s_j^{\prime} (\lambda, 0) - 1 = 0$ and by $c_j (\lambda, x_j)$ the
solution which satisfies the conditions $c_j (\lambda, 0) - 1 = c_j^{\prime} (\lambda, 0) = 0$. Then the {\it characteristic
function}, i.e. an entire function whose zeros coincide with the spectrum of the problem
can be expressed via $s_j (\lambda, l_j )$,  $s^{\prime}_j (\lambda, l_j )$, $c_j (\lambda, l_j )$ and  $c^{\prime}_j (\lambda, l_j )$. To do it we introduce
the following system of vector functions $\psi_j (\lambda, \vec{x}) =  \{0, 0, . . . , c_j(\lambda, x_j), . . . , 0\}^T$ and
$\psi_{j+n}(\lambda, \vec{x}) = \{0, 0, . . . , s_j (\lambda, x_j), . . . , 0\}^T$ for $j = 1, 2, . . . , g$, where $g$ is the number
of edges in the graph, $\vec{x}=\{x_1,x_2,...,x_g\}^T$. As in \cite{PP}, we denote by $L_j$ $(j = 1, 2, . . . , 2g)$ the linear functionals generated by
(\ref{2.2})--(\ref{2.8}). Then $\Phi (\lambda) = (L_j(\psi_k(\lambda, \vec{l})))^{2g}_{j,k}$ where $\vec{l}=\{l_1,l_2,...,l_g\}^T$ is the characteristic matrix which represents the
system of linear equations describing the continuity and Kirchhoff conditions for the interior
vertices. Then we call
$$
\phi_N(\lambda) := \det(\Phi(\lambda))
$$
the {\it characteristic function} of problem (\ref{2.2})--(\ref{2.8}).  
We are interested also in the problem generated by the same equations and all 
boundary and matching conditions the same, but with the condition
\begin{equation}
\label{2.9}
y_j (0 ) = 0 
\end{equation}
instead of (\ref{2.8}) at ${\bf v}$. We denote the characteristic function of problem (\ref{2.2})--(\ref{2.7}), (\ref{2.9}),
 by $\Phi_D(\lambda)$. We  call  (\ref{2.7}), (\ref{2.9}) the  {\it generalized Dirichlet condition}.  In case when {\bf v} is a pendant vertex, conditions (\ref {2.7}), (\ref{2.9}) are reduced to the usual  Dirichlet boundary
condition. 
Let us assume that the graph $G$ is separable, i.e of connectivity 1   and let the  root {\bf v} be the cut-vertex. We divide our graph $G$ into two subgraphs
$G_1$ and $G_2$ having the only  common  vertex {\bf v}. Denote by $\Phi^j_N(\lambda)$ ($j=1,2$) the characteristic function corresponding to subgraph $G_j$ with  Neumann condition at ${\bf v}$ and by $\Phi^j_D(\lambda)$ the characteristic function corresponding to subgraph $G_j$ with  Dirichlet condition at ${\bf v}$. 

Formula (\ref{1.1}) describes series connection of subgraphs.
In order to describe the characteristic functions for series and parallel connected subgraphs we need to generalize the notions of $\Phi_{DD}$, $\Phi_{DN}$, $\Phi_{ND}$ and  $\Phi_{NN}$ introduced in the beginning of this Section for boundary value problems on an interval.

Let the vertices ${\bf v}_{in}$ and ${\bf v}_{out}$ be pendant vertices and let  the edge $e_1$ be outgoing away from  ${\bf v}_{in}$ and the edge $e_g$ be  incoming to  ${\bf v}_{out}$. Denote by $w={\bf v}_{out}$ the second vertex
incident with $e_g$. Let $w$ have no incoming edges and denote the outgoing from $w$ edges by $e_{g-d(w)+1},..,e_{g}$.

\begin{figure}[tbh]
  \begin{center}
    \includegraphics{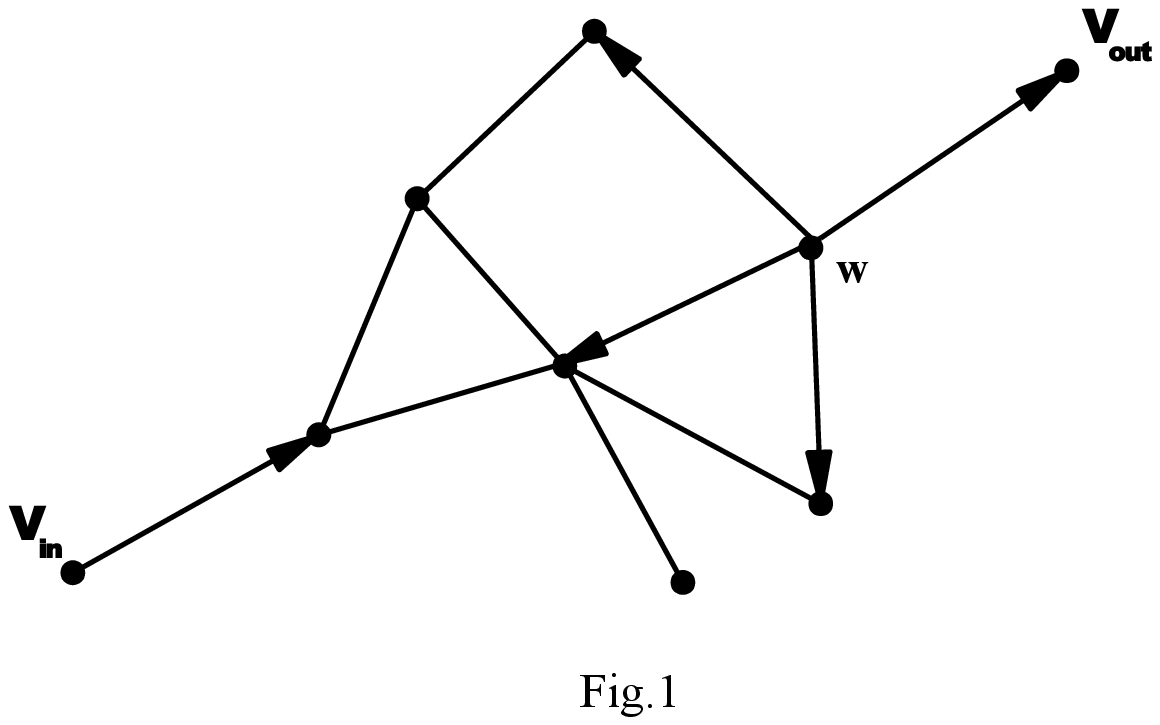}
  \end{center}
\end{figure}

Let us consider the Dirichlet-Dirichlet problem  (\ref{2.2})--(\ref{2.7}), (\ref{2.9}) which in our terms is as follows:
\begin{equation}
\label{2.10}
y_1(0)=0
\end{equation}
at ${\bf v}_{in}$,
\begin{equation}
\label{2.11}
y^{\prime}_
j (0) + \beta_j y_j (0) = 0, 
\end{equation}
or
\begin{equation}
\label{2.12}
y^{\prime}_j (l_j) + \beta_j y_j (l_j) = 0, 
\end{equation}
at each pendant vertex except of ${\bf v}_{in}$ and ${\bf v}_{out}$.
 For each  interior vertex which is not  $w$  with
incoming edges $e_j$ and outgoing edge $e_k$ the continuity conditions are
\begin{equation}
\label{2.13}
y_j (l_j ) = y_k(0), 
\end{equation}
and  Kirchhoff's condition is
\begin{equation}
\label{2.14}
\sum\limits_ky_k^{\prime}
(0) =\sum\limits_jy_j^{\prime}(l_j).
\end{equation}
 At $w$ we have 
\begin{equation}
\label{2.15}
y_{g-d(w)+1}(0)=y_{g-d(w)+2}(0)=...=y_{g-1}(0)=y_g(0),
\end{equation}
and
\begin{equation}
\label{2.16}
y^{\prime}_{g}(0)=-\sum\limits_{k=g-d(w)+1}^{g-1}y^{\prime}_k(0)
\end{equation}
and
\begin{equation}
\label{2.17}
y_g(l_g)=0
\end{equation}
at ${\bf v}_{out}$.

We substitute
$$
y_i=B_ic_i(\lambda,x_i)+A_is_i(\lambda,x_i)
$$
into (\ref{2.10})--(\ref{2.17}) and obtain 
\begin{equation}
\label{2.18}
B_1=0,
\end{equation} 
\begin{equation}
\label{2.19}
A_j+\beta_jB_j=0,
\end{equation}
\begin{equation}
\label{2.20}
B_jc_j^{\prime}(\lambda,l_j)+A_js_j^{\prime}(\lambda,l_j)+\beta_j(B_jc_j(\lambda,l_j)+A_js_j(\lambda,l_j))=0.
\end{equation}
 We can also write the continuity and Kirchhoff's condition at  ${\tilde v}$ as
\begin{equation}
\label{2.21}
B_{g-d(w)+1}=B_{g-d(w)+2}=...=B_{g-1}
\end{equation}
\begin{equation}
\label{2.22} 
B_{g-1}=B_g,
\end{equation}
\begin{equation}
\label{2.23}
A_g=-\sum\limits_{k=g-d(w)+1}^{g-1}A_k,
\end{equation}
\begin{equation}
\label{2.24}
B_gc_g(\lambda,l_g)+A_gs_g(\lambda,l_g)=0.
\end{equation}
These equations can be written in a matrix form:
\begin{equation}
\label{2.25}
\left(\begin{array}{cccccccccccc}
1&0&...&...&...&0       &     0&0&0&...&...&0\\
...&\textcolor{red}{...}&\textcolor{red}{...}&\textcolor{red}{...}&\textcolor{red}{...}&...& \times &\textcolor{red}{...}&\textcolor{red}{...}&\textcolor{red}{...}&\textcolor{red}{...}&...\\
...&\textcolor{red}{...}&\textcolor{red}{...}&\textcolor{red}{...}&\textcolor{red}{...}&...& \times &\textcolor{red}{...}&\textcolor{red}{...}&\textcolor{red}{...}&\textcolor{red}{...}&...\\
...&\textcolor{red}{...}&\textcolor{red}{...}&\textcolor{red}{...}&\textcolor{red}{...}&...& \times &\textcolor{red}{...}&\textcolor{red}{...}&\textcolor{red}{...}&\textcolor{red}{...}&...\\
...&\textcolor{red}{...}&\textcolor{red}{...}&\textcolor{red}{...}&\textcolor{red}{...}&...& \times &\textcolor{red}{...}&\textcolor{red}{...}&\textcolor{red}{...}&\textcolor{red}{...}&...\\
0&0&...&0& 1 &-1&0&0&...&...& ... &0\\
...&\textcolor{red}{...}&\textcolor{red}{...}&\textcolor{red}{...}&\textcolor{red}{...}&...& \times &\textcolor{red}{...}&\textcolor{red}{...}&\textcolor{red}{...}&\textcolor{red}{...}&...\\
...&\textcolor{red}{...}&\textcolor{red}{...}&\textcolor{red}{...}&\textcolor{red}{...}&...& \times &\textcolor{red}{...}&\textcolor{red}{...}&\textcolor{red}{...}&\textcolor{red}{...}&...\\
...&\textcolor{red}{...}&\textcolor{red}{...}&\textcolor{red}{...}&\textcolor{red}{...}&...& \times &\textcolor{red}{...}&\textcolor{red}{...}&\textcolor{red}{...}&\textcolor{red}{...}&...\\
...&\textcolor{red}{...}&\textcolor{red}{...}&\textcolor{red}{...}&\textcolor{red}{...}&...& \times &\textcolor{red}{...}&\textcolor{red}{...}&\textcolor{red}{...}&\textcolor{red}{...}&...\\
0&0&...&...& ... &0&0&0&...& 1 & 1 &1\\
0&0&...&...&0&c_g&0&...&...&...&0&s_g
\end{array}\right) \left(\begin{array}{c}
B_1\\ B_2 \\ .\\.\\B_{g-1}\\B_g \\A_1 \\A_2 \\ ,\\.\\A_{g-1}\\A_g
\end{array}\right)=0.
\end{equation}
Here $s_j=s_j(\lambda, l_j)$, $s_j^{\prime}=s_j^{\prime}(\lambda, l_j)$, $c_j=c_j(\lambda, l_j)$, $c_j^{\prime}=c^{\prime}_j(\lambda, l_j)$.

The determinant of the matrix in (\ref{2.24}) is $\Phi_{DD}(\lambda)$.

Let us delete  equations (\ref{2.18}), (\ref{2.22}), (\ref{2.23}) and (\ref{2.24}) from (\ref{2.25}) and set $B_1=B_g=A_g=0$ and $A_1=1$. Then we obtain a nonhomogeneus $(g-4)$x$(g-4)$ system of linear algebraic equation with respect to unknowns
 $B_2, B_3,...,B_{g-1}, A_2, A_3, ..., A_{g-1}$:

\begin{equation}
\label{2.26}
\left(\begin{array}{cccccccc}
...&...&...&...&...&...&...&...\\
...&...&...&...&...&...&...&...\\
...&...&...&...&...&...&...&...\\
...&...&...&...&...&...&...&...\\
...&...&...&...&...&...&...&...\\
...&...&...&...&...&...&...&...\\
...&...&...&...&...&...&...&...\\
...&...&...&...&...&...&...&...\\
\end{array}\right) \left(\begin{array}{c}
\ B_2 \\ .\\.\\B_{g-1} \\A_2 \\ .\\.\\A_{g-1}
\end{array}\right)=-\left(\begin{array}{c} \times \\ \times \\ \times \\ \times \\ \times \\ \times \\ \times \\ \times \\\end{array}\right),
\end{equation}

Denote the determinant of the matrix of the obtained system by $\Delta(\lambda)$. Solving this system we find 
\begin{equation}
\label{2.27}
B_{g-1}^D=\frac{\Delta_{B_{g-1}}^D(\lambda)}{\Delta(\lambda)}, \ \ \ A_{g-k}^D=\frac{\Delta_{A_{g-k}}^D(\lambda)}{\Delta(\lambda)}
\ \ \ k=1,2,..., d(w)-1,
\end{equation}
where $\Delta_{B_{g-1}}^D$ and $\Delta_{A_{k}}^D$ are the corresponding cofactors:
$$
\Delta_{B_{g-1}}^D(\lambda)=det\left|\begin{array}{cccccccc}
...&...&...& \times &...&...&...&...\\
...&...&...& \times &...&...&...&...\\
...&...&...& \times &...&...&...&...\\
...&...&...& \times &...&...&...&...\\
...&...&...& \times &...&...&...&...\\
...&...&...& \times &...&...&...&...\\
...&...&...& \times &...&...&...&...\\
...&...&...& \times &...&...&...&...\\
\end{array}\right|,
$$

$$
\Delta_{A_{g-1}}^D(\lambda)=det\left|\begin{array}{cccccccc}
...&...&...&...&...&...&...& \times \\
...&...&...&...&...&...&...& \times \\
...&...&...&...&...&...&...& \times \\
...&...&...&...&...&...&...& \times \\
...&...&...&...&...&...&...& \times \\
...&...&...&...&...&...&...& \times \\
...&...&...&...&...&...&...& \times \\
...&...&...&...&...&...&...& \times \\
\end{array}\right|,
\ \
\Delta_{A_{g-2}}^D(\lambda)=det\left|\begin{array}{cccccccc}
...&...&...&...&...&...& \times &...\\
...&...&...&...&...&...& \times &...\\
...&...&...&...&...&...& \times &...\\
...&...&...&...&...&...& \times &...\\
...&...&...&...&...&...& \times &...\\
...&...&...&...&...&...& \times &...\\
...&...&...&...&...&...& \times &...\\
...&...&...&...&...&...& \times &...\\
\end{array}\right|
$$
and so on up to $\Delta^D_{A_{g- d(w)+1}}$.

We use upper index $D$  to underline that 
the formulae correspond to the case of Dirichlet boundary condition at ${\bf v}_{in}$. Notice that $\Delta$ does not depend on the condition at ${\bf v}_{in}$ while it is the determinant of the matrix obtained from the matrix of (\ref{2.27}) by deleting among others the first column and the $(g+1)$-th column. 

On the other hand, it is easy to notice that these cofactors appear in the expansion of the determinant 
\begin{equation}
\label{2.28}
\Phi_{DD}(\lambda)=c_g(\lambda,l_g)\tilde{\Delta}_{B_{g-1}}^D(\lambda)+s_g(\lambda,l_g)\sum\limits_{k=g-d(w)+1}^{g-1}\tilde{\Delta}_{A_k}^D(\lambda)
\end{equation} 
because it is easy to see that $\tilde{\Delta}_{B_{g-1}}^D(\lambda)=\Delta_{B_{g-1}}^D(\lambda)$ and $\tilde{\Delta}_{A_{k}}^D(\lambda)=\Delta_{A_{k}}^D(\lambda)$. 

In the same way we arrive at
\begin{equation}
\label{2.29}
\Phi_{DN}(\lambda)=c^{\prime}_g(\lambda,l_g)\Delta_{B_{g-1}}^D(\lambda)+s^{\prime}_g(\lambda,l_g)\sum\limits_{k=g-d(w)+1}^{g-1}\Delta_{A_k}^D(\lambda).
\end{equation}
Now we consider the same problem but with Neumann condition 
$y_1^{\prime}(0)=0$ at ${\bf v}_{in}$. Then we obtain $A_1=0$ instead of (\ref{2.18}) and
$$
\left(\begin{array}{cccccccccccc}
0&0&...&...&...&0 & 1&0&0&...&...&0\\
...&...&...&...&...&...&...&...&...&...&...&...\\
...&...&...&...&...&...&...&...&...&...&...&...\\
...&...&...&...&...&...&...&...&...&...&...&...\\
...&
...&...&...&...&...&...&...&...&...&...\\
0&0&...&0&1&-1&0&0&...&...&...&0\\
...&...&...&...&...&...&...&...&...&...&...&...\\
...&...&...&...&...&...&...&...&...&...&...&...\\
...&...&...&...&...&...&...&...&...&...&...&...\\
...&...&...&...&...&...&...&...&...&...&...&...\\
0&0&...&...& 0 &0&0&0&...& 1 & 1 &-1\\
0&0&...&...&0&c_g&0&...&...&...&0&s_g

\end{array}\right) \left(\begin{array}{c}
B_1\\ B_2 \\ .\\.\\B_{g-1}\\B_g \\A_1 \\A_2 \\ ,\\.\\A_{g-1}\\A_g
\end{array}\right)=0
$$
instead of (\ref{2.25}). In the same way as (\ref{2.27}) we obtain
\begin{equation}
\label{2.30}
B_{g-1}^N=\frac{\Delta_{B_{g-1}}^N(\lambda)}{\Delta(\lambda)}, \ \ \ A_{g-k}^N=\frac{\Delta_{A_{g-k}}^N(\lambda)}{\Delta(\lambda)}
\ \ \ k=1,2,..., d(w)-1,
\end{equation}
where $\Delta_{B_{g-1}}^N$ and $\Delta_{A_{k}}^N$ are the corresponding cofactors. Upper index $N$ we use to underline that 
the formulae correspond to the case of Neumann boundary condition at ${\bf v}_{in}$. Also we have
$$
\Phi_{ND}(\lambda)=c_g(\lambda,l_g)\Delta_{B_{g-1}}^N(\lambda)+s_g(\lambda,l_g)\sum\limits_{k=g-d(w)+1}^{g-1}\Delta_{A_k}^N(\lambda),
$$
$$
\Phi_{NN}(\lambda)=c^{\prime}_g(\lambda,l_g)\Delta_{B_{g-1}}^N(\lambda)+s^{\prime}_g(\lambda,l_g)\sum\limits_{k=g-d(w)+1}^{g-1}\Delta_{A_k}^N(\lambda).
$$

\section{ Series connection}
\setcounter{equation}{0} \hskip0.25in

Let us consider two connected graphs $G^j$ ($j=1,2$). We choose two vertices  ${\bf v}_{in}^j$ and ${\bf v}_{out}^j$ in each of them as the entrance and exit vertices. In this Section we investigate series connection of these graphs. We denote by $\Phi^j_{NN}$ the characteristic function of the boundary value problem on the graph $G_j$ with Neumann boundary conditions at 
$\bf{v}_{in}^j$ and $\bf{v}_{out}^j$. At all the other interior vertices generalized Neumann conditions (continuity and Kirchhoff conditions) are imposed while any self-adjoint conditions of the form (\ref{2.3}) or (\ref{2.4}) are  imposed at pendant vertices. In the same way, we 
denote by $\Phi^j_{ND}$ the characteristic function of the boundary value problem on the graph $G_j$ with Neumann boundary condition at 
$\bf{v}_{in}^j$ and Dirichlet boundary condition at $\bf{v}_{out}^j$,  by $\Phi^j_{DN}$ the characteristic function of the boundary value problem on the graph $G_j$ with Dirichlet boundary condition at 
$\bf{v}_{in}^j$ and Neumann boundary condition at $\bf{v}_{out}^j$,  by $\Phi^j_{DD}$ the characteristic function of the boundary value problem on the graph $G_j$ with Dirichlet boundary condition at 
$\bf{v}_{in}^j$ and at $\bf{v}_{out}^j$.

If we connect ${\bf v}_{out}^1$ with ${\bf v}_{in}^2$ then we obtain a new graph $G=G_1\cup G_2$ with a cut-vertex 
${\bf v}_{out}^1={\bf v}_{in}^2$. Let us denote by $\Phi_{NN}(\lambda)$, $\Phi_{ND}(\lambda)$, $\Phi_{DN}(\lambda)$ and $\Phi_{DD}(\lambda)$ the characteristic functions of the problems on $G$ with Neumann conditions at ${\bf v}_{in}^1$ and ${\bf v}_{out}^2$, with Neumann conditions at ${\bf v}_{in}^1$ and Dirichlet at ${\bf v}_{out}^2$, with Dirichlet condition at ${\bf v}_{in}^1$ and Neumann at ${\bf v}_{out}^2$ and with Dirichlet conditions at ${\bf v}_{in}^1$ and ${\bf v}_{out}^2$. 

It is clear from (\ref{1.1}) that
\begin{equation}
\label{3.1}
\Phi_{NN}(\lambda)=\Phi_{NN}^1(\lambda)\Phi_{DN}^2(\lambda)+\Phi_{ND}^1(\lambda)\Phi_{NN}^2(\lambda),
\end{equation}
\begin{equation}
\label{3.2}
\Phi_{ND}(\lambda)=\Phi_{NN}^1(\lambda)\Phi_{DD}^2(\lambda)+\Phi_{ND}^1(\lambda)\Phi_{ND}^2(\lambda),
\end{equation}
\begin{equation}
\label{3.3}
\Phi_{DN}(\lambda)=\Phi_{DN}^1(\lambda)\Phi_{DD}^2(\lambda)+\Phi_{DD}^1(\lambda)\Phi_{NN}^2(\lambda),
\end{equation}
\begin{equation}
\label{3.4}
\Phi_{DD}(\lambda)=\Phi_{DN}^1(\lambda)\Phi_{DD}^2(\lambda)+\Phi_{DD}^1(\lambda)\Phi_{ND}^2(\lambda).
\end{equation}
Using these identities we obtain an analogue  of the Lagrange identity:
\begin{equation}
\label{3.5}
\Phi_{ND}(\lambda)\Phi_{DN}(\lambda)-\Phi_{NN}(\lambda)\Phi_{DD}(\lambda)=
\end{equation}
$$
(\Phi_{ND}^1(\lambda)\Phi_{DN}^1(\lambda)-\Phi_{NN}^1(\lambda)\Phi_{DD}^1(\lambda))(\Phi_{ND}^2(\lambda)\Phi_{DN}^2(\lambda)-\Phi_{NN}^2(\lambda)\Phi_{DD}^2(\lambda)).
$$

\section{Auxiliary results}
\setcounter{equation}{0} \hskip0.25in

Denote by $s_{1,j}(\lambda,x_{1,j})$ the solution of the equation
$$
-y_{1,j}^{\prime\prime}+q_{1,j}(x_{1,j})y_{1,j}=\lambda^2y_{1,j},  \ \  \  j=1,2
$$
on the edge of $G_j$ incident with ${\bf v^j_{in}}$ which satisfies $s_{1,j}(\lambda,0)=s_{1,j}^{\prime}(\lambda,0)-1=0$ and by 
 $c_{1,j}(\lambda,x_{1,j})$ the solution  which satisfies $c_{1,j}(\lambda,0)-1=c_{1,j}^{\prime}(\lambda,0)=0$. 

Let us consider a boundary value problem on $G_1$ which consists of equations 
\begin{equation}
\label{4.1}
-y_{i,1}^{\prime\prime}+q_{i,1}(x_{i,1})y_{i,1}=\lambda^2y_{i,1}, \ \ \  i=1,2,..., g_1,
\end{equation}
continuity and Kirchhoff conditions at all interior vertices, conditions $y_{1,1}(0)=y_{1,1}^{\prime}(0)-1=0$ at ${\bf v}^1_{in}$ and no condition at ${\bf v}_{out}^1$  There exists a solution of this problem, maybe not unique, of the form  
$Y_{1}(\lambda,\vec{x}_1)=(s_{1,1}(\lambda,x_{1,1}), y_{2,1}(\lambda,x_{2,1}), y_{3,1}(\lambda,x_{3,1}), ..., y_{g_1,1}(\lambda,x_{g_1,1}))^T$, where $\vec{x}_1=\{x_{1,1}, x_{2,1}, ..., x_{g,1}\}^T$ is the coordinate vector corresponding to $G_1$
In the same way we define the solution of the problem generated by equations (\ref{4.1}),  conditions  $y_{1,1}(0)-1=y_{1,1}^{\prime}(0)=0$ at ${\bf v}^1_{in}$ and no condition at ${\bf v}_{out}^1$: 
$U_{1}(\lambda,\vec{x}_1)=(c_{1,1}(\lambda,x_{1,1}), u_{2,1}(\lambda,x_{2,1}), u_{3,1}(\lambda,x_{3,1}), ..., u_{g_1,1}(\lambda,x_{g_1,1}))^T$.

If we substitute $Y_1(\vec{x}_1)$ into equations (\ref{2.10})--(\ref{2.16}) written for $G_1$,  we obtain 
\begin{equation}
\label{4.2}
y_{g_1,1}(\lambda,x_{g_1,1})=\frac{\Delta_{B_{g_1-1,1}}^D(\lambda)}{\Delta_{1}(\lambda)}c_{g_1,1}(\lambda,x_{g_1,1})+\sum\limits_{k=g-d(w)+1}^{g-1}\frac{\Delta_{A_{k,1}}^D(\lambda)}{\Delta_{1}(\lambda)}s_{g_1,1}.(\lambda,x_{g_1,1})
\end{equation}
In the same way
\begin{equation}
\label{4.3}
u_{g_1,1}(\lambda,x_{g_1,1})=\frac{\Delta_{B_{g_1-1,1}}^N(\lambda)}{\Delta_{1}(\lambda)}c_{g_1,1}(\lambda,x_{g_1,1})+\sum\limits_{k=g-d(w)+1}^{g-1}\frac{\Delta_{A_{k,1}}^N(\lambda)}{\Delta_{1}(\lambda)}s_{g_1,1}(\lambda,x_{g_1,1}).
\end{equation}
In the same way we define the corresponding solutions for $G_2$:
$Y_{2}(\lambda,\vec{x}_2)=(s_{1,2}(\lambda,x_{1,2}), y_{2,1}(\lambda,x_{2,2}), y_{3,1}(\lambda,x_{3,2}), ..., y_{g_1,2}(\lambda,x_{g_1,2}))^T$ which satisfies equations 
\begin{equation}
\label{4.4}
-y_{i,2}^{\prime\prime}+q_{i,2}(x_{i,2})y_{i,2}=\lambda^2y_{i,2}, \ \ \ i=1,2,..., g_2,
\end{equation}
 and conditions  $y_{1,2}(0)=y_{1,2}^{\prime}(0)-1=0$ at ${\bf v}^2_{in}$ and no condition at ${\bf v}_{out}^2$ and 
$U_{2}(\lambda,\vec{x}_2)=(c_{1,2}(\lambda,x_{1,2}), u_{2,2}(\lambda,x_{2,2}), u_{3,2}(\lambda,x_{3,2}), ..., u_{g_2,2}(\lambda,x_{g_2,2}))^T$ which satisfies  $u_{1,2}(0)-1=u_{1,2}^{\prime}(0)=0$ at ${\bf v}^2_{in}$ and no condition at ${\bf v}_{out}^2$.  Then
\begin{equation}
\label{4.5}
y_{g_2,2}(\lambda,x_{g_2,2})=\frac{\Delta_{B_{g_2-1,2}}^D(\lambda)}{\Delta_{2}(\lambda)}c_{g_2,2}(\lambda,x_{g_2,2})+\sum\limits_{k=g-d(w)+1}^{g-1}\frac{\Delta_{A_{k,2}}^D(\lambda)}{\Delta_{2}(\lambda)}s_{g_2,2}(\lambda,x_{g_2,2}),
\end{equation}
\begin{equation}
\label{4.6}
u_{g,2}(\lambda,x_{g_2,2})=\frac{\Delta_{B_{g_2-1,2}}^N(\lambda)}{\Delta_{2}(\lambda)}c_{g_2,2}(\lambda,x_{g_2,2})+\sum\limits_{k=g-d(w)+1}^{g-1}\frac{\Delta_{A_{k,2}}^N(\lambda)}{\Delta_{2}(\lambda)}s_{g_2,2}(\lambda,x_{g_2,2}).
\end{equation}

\section{Parallel connection}
\setcounter{equation}{0} \hskip0.25in

Now we consider parallel connection of $G_1$ with $G_2$, i.e. a graph $G$ obtained by connection ${\bf v}_{in}^1$ with ${\bf v}_{in}^2$ and ${\bf v}_{out}^1$ with ${\bf v}_{out}^2$. To simplify situation let us assume that ${\bf v}^j_{in}$ and ${\bf v}^j_{out}$ are pendant  vertices of $G_j$ for $j=1,2$, respectively, (see Fig. 2). 

\begin{figure}[tbh]
  \begin{center}
    \includegraphics{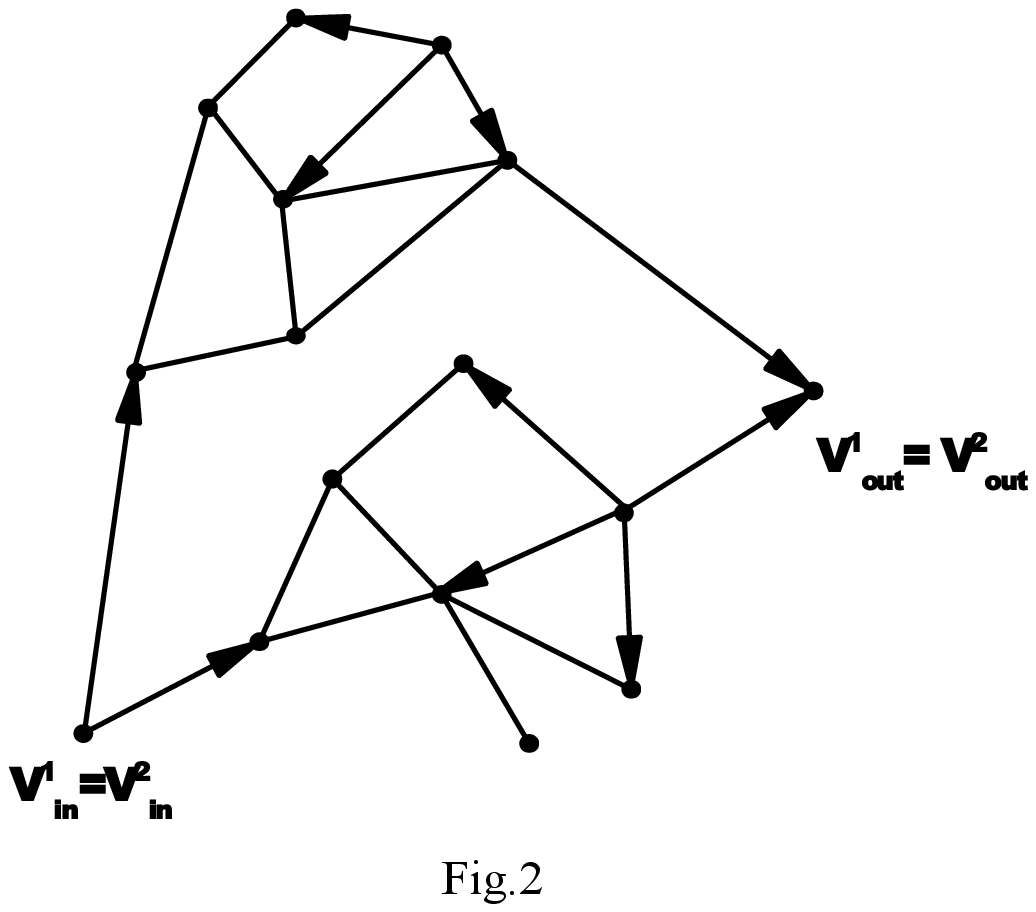}
  \end{center}
\end{figure}

 Let us denote by $\Phi_{NN}(\lambda)$, $\Phi_{ND}(\lambda)$, $\Phi_{DN}(\lambda)$ and $\Phi_{DD}(\lambda)$ the characteristic functions of the problems on $G$ with Neumann conditions at ${\bf v}_{in}^1={\bf v}_{in}^2$ and ${\bf v}_{out}^1={\bf v}_{out}^2$, with Neumann conditions at ${\bf v}_{in}^1={\bf v}_{in}^2$  and Dirichlet at ${\bf v}_{out}^1={\bf v}_{out}^2$, with Dirichlet condition at ${\bf v }_{in}^1={\bf v}_{in}^2$ and Neumann at ${\bf v}_{out}^1={\bf v}_{out}^2$  and with Dirichlet conditions at ${\bf v}_{in}^1={\bf v}_{in}^2$  and at ${\bf v}_{out}^1={\bf v}_{out}^2$, respectively. 

It is clear that 
\begin{equation}
\label{5.i}
\Phi_{DD}(\lambda)=\Phi_{DD}^1(\lambda)\Phi_{DD}^2(\lambda),
\end{equation}
\begin{equation}
\label{5.ii}
\Phi_{DN}(\lambda)=\Phi_{DN}^1(\lambda)\Phi_{DD}^2(\lambda)+\Phi_{DD}^1(\lambda)\Phi_{DN}^2(\lambda),
\end{equation}
\begin{equation}
\label{5.iii}
\Phi_{ND}(\lambda)=\Phi_{ND}^1(\lambda)\Phi_{DD}^2(\lambda)+\Phi_{DD}^1(\lambda)\Phi_{ND}^2(\lambda).
\end{equation}
Our aim is to express $\Phi_{NN}(\lambda)$ via $\Phi_{NN}^j(\lambda)$, $\Phi_{ND}^j(\lambda)$, $\Phi_{DN}^j(\lambda)$ and $\Phi_{DD}^j(\lambda)$ ($j=1,2$).

Let us look for solution of Neumann-Neumann problem on $G$ in the form
$\{R_{11}Y_1+R_{21}U_1, R_{12}Y_2+R_{22}U_2\}$, where $R_{ij}$ are constants. Then the continuity condition at ${\bf v}_{in}^1={\bf v}_{in}^2$ is
\begin{equation}
\label{5.4}
R_{21}=R_{22}.
\end{equation}
The Kirchhoff condition at ${\bf v}_{in}^1={\bf v}_{in}^2$ is 
\begin{equation}
\label{5.5}
R_{11}= - R_{12}.
\end{equation}
Continuity condition at ${\bf v}_{out}^1={\bf v}_{out}^2$ is
\begin{equation}
\label{5.6}
R_{11}y_{g_1,1}(\lambda,l_{g_1,1})+R_{21}u_{g_1,1}(\lambda, l_{g_1,1})=R_{12}y_{g_2,2}(\lambda,l_{g_2,2})+R_{22}u_{g_2,2}(\lambda, l_{g_2,2}).
\end{equation}
The Kirchhoff condition at ${\bf v}_{out}^1={\bf v}_{out}^2$ is 
\begin{equation}
\label{5.7}
R_{11}y^{\prime}_{g_1,1}(\lambda,l_{g_1,1})+R_{21}u_{g_1,1}^{\prime}(\lambda, l_{g_1,1})+R_{12}y^{\prime}_{g_2,2}(\lambda,l_{g_2,2})+R_{22}u^{\prime}_{g_2,2}(\lambda, l_{g_2,2})=0.
\end{equation}

It is clear that if $y_{g_1,1}(\lambda, l_{g_1,1})=0$, then $Y_1$ is an eigenvector of the problem on $G_1$ with  Dirichlet conditions at ${\bf v}^1_{in}$ and at  ${\bf v}^1_{out}$. It means that the set of zeros of $y_{g_1,1}(\lambda, l_{g_1,1})$ coincides with the set of zeros of $\Phi_{DD}^1(\lambda)$. 

Using (\ref{2.29}), (\ref{2.30}) we obtain from (\ref{4.2}), (\ref{4.3}), (\ref{4.5}) and (\ref{4.6})  that
 $y_{g_1,1}(\lambda, l_{g_1,1})=\frac{\Phi_{DD}^1(\lambda)}{\Delta_1(\lambda)}$,  $u_{g_1,1}(\lambda, l_{g_1,1})=\frac{\Phi_{ND}^1(\lambda)}{\Delta_1(\lambda)}$,
  $y_{g_2,2}(\lambda, l_{g_2,2})=\frac{\Phi_{DD}^2(\lambda)}{\Delta_2(\lambda)}$, 
 $u_{g_2,2}(\lambda, l_{g_2,2})=\frac{\Phi_{ND}^2(\lambda)}{\Delta_2(\lambda)}$, 
 $y^{\prime}_{g_1,1}(\lambda, l_{g_1,1})=\frac{\Phi_{ND}^1(\lambda)}{\Delta_1(\lambda)}$,  
$u^{\prime}_{g_1,1}(\lambda, l_{g_1,1})=\frac{\Phi_{NN}^1(\lambda)}{\Delta_1(\lambda)}$,  
$y^{\prime}_{g_2,2}(\lambda, l_{g_2,2})=\frac{\Phi_{DN}^2(\lambda)}{\Delta_2(\lambda)}$ and  $u^{\prime}_{g_2,2}(\lambda, l_{g_2,2})=\frac{\Phi_{NN}^2(\lambda)}{\Delta_2(\lambda)}$.

The determinant of system (\ref{5.4})--(\ref{5.7}) is
$$
D(\lambda)\mathop{=}\limits^{def}\det \left|\begin{array}{cccc}
0&0&1&-1\\
1&1&0&0\\
\frac{\Phi^1_{DD}(\lambda)}{\Delta_1(\lambda)}&-\frac{\Phi^2_{DD}(\lambda)}{\Delta_2(\lambda)}&\frac{\Phi_{ND}^1(\lambda)}{\Delta_1(\lambda)}&-\frac{\Phi^2_{ND}(\lambda)}{\Delta_2(\lambda)}\\
\frac{\Phi^1_{DN}(\lambda)}{\Delta_1(\lambda)}&\frac{\Phi^2_{DN}(\lambda)}{\Delta(\lambda)}&\frac{\Phi_{NN}^1(\lambda^2)}{\Delta_2(\lambda)}&\frac{\Phi^2_{NN}(\lambda)}{\Delta_2(\lambda)}
\end{array}\right|=
$$
\begin{equation}
\label{5.8}
\left(\frac{\Phi_{DD}^1(\lambda)}{\Delta_1(\lambda)}+\frac{\Phi_{DD}^2(\lambda)}{\Delta_2(\lambda)}\right)\left(\frac{\Phi_{NN}^1(\lambda)}{\Delta_1(\lambda)}+\frac{\Phi_{NN}^2(\lambda)}{\Delta_2(\lambda)}\right)-
\end{equation}
$$
\left(\frac{\Phi_{DN}^1(\lambda)}{\Delta_1(\lambda)}-\frac{\Phi_{DN}^2(\lambda)}{\Delta_2(\lambda)}\right)\left(\frac{\Phi_{ND}^1(\lambda)}{\Delta_1(\lambda)}-\frac{\Phi_{ND}^2(\lambda)}{\Delta_2(\lambda)}\right).
$$

It is clear that to obtain the characteristic function $\Phi_{NN}(\lambda)$ the determinant (\ref{5.5}) must be multiplied by $(\Delta_1(\lambda)\Delta_2(\lambda))^p$ (with some $p\geq  1$) to be an entire function. Let us show that this $p$ must be $1$. 

To prove it we notice that $\Phi^j_{DD}(\lambda)$ is an entire function of exponential type $L^{(j)}=\mathop{\sum}\limits_{k=1}^{g_j}l_k^{(j)}$, where $l_k^{(j)}$ are the lengths of the edges of the subgraph $G_j$ and $g_j$ is their number. The function $\Delta_j(\lambda)$ is an entire function of exponential type $\mathop{\sum}\limits_{k=2}^{g_j-1}l_k^{(j)}=L^{(j)}-l_1^{(j)}-l_{g_j}^{(j)}$.

The graph $G_j$ consists of two edges $e_1$ and $e_{g_j}$ and the subgraph  $G_j^0$
(see Fig. 2) which are series connected. Then for $e_1^{(j)}$ and $e_{g_j}^{(j)}$ we have
\begin{equation}
\label{5.9}
c_{1,j}(\lambda,l_1^{(j)})s_{1,j}^{\prime}(\lambda, l_1^{(j)})-c_{1,j}^{\prime}(\lambda, l_1^{(j)})s_{1,j}(\lambda,l_1^{(j)})=1,
\end{equation}
\begin{equation}
\label{5.10}
c_{g_j,j}(\lambda,l_{g_j}^{(j)})s_{g_j,j}^{\prime}(\lambda, l_{g_j}^{(j)})-c_{g_j,j}^{\prime}(\lambda, l_{g_j}^{(j)})s_{g_j,j}(\lambda,l_{g_j}^{(j)})=1.
\end{equation}
Applying the analogue of Lagrange identity for series connected subgraphs (\ref{3.5}) and using (\ref{5.6}) and (\ref{5.7}) we obtain
$$
\Phi_{ND}^{(j)}(\lambda)\Phi_{DN}^{(j)}(\lambda)-\Phi_{NN}^{(j)}(\lambda)\Phi_{DD}^{(j)}(\lambda)=
$$
$$
\Phi_{ND}^{j0}(\lambda)\Phi_{DN}^{j0}(\lambda)-\Phi_{NN}^{j0}(\lambda)\Phi_{DD}^{j0}(\lambda),
$$
where quantities with the zero upper index correspond to $G_j^0$. Thus $\Phi_{ND}^{(j)}(\lambda)\Phi_{DN}^{(j)}(\lambda)-\Phi_{NN}^{(j)}(\lambda)\Phi_{DD}^{(j)}(\lambda)$ is an entire function of exponential type $L^{(j)}-l_{g_j}^{(j)}-l_{n_j}^{(j)}$. Taking into account that $\Delta_j(-\lambda)=\Delta_j(\lambda)$ and $\Delta_j(\overline{\lambda})=\overline{\Delta_j(\lambda)}$  we conclude that
\begin{equation}
\label{5.11}
\frac{\Phi_{ND}^{(j)}(\lambda)\Phi_{DN}^{(j)}(\lambda)-\Phi_{NN}^{(j)}(\lambda)\Phi_{DD}^{(j)}(\lambda)}{\Delta_j(\lambda)}\rightarrow C,  \   \  \  \ |Im \ \lambda|\rightarrow \infty 
\end{equation} 
where $C$ is a real constant. Therefore, using (\ref{5.5}) we obtain
\begin{equation}
\label{5.12}
\Delta_1(\lambda)\Delta_2(\lambda)D(\lambda)=
\end{equation}
$$
\frac{\Phi^1_{DD}(\lambda)\Phi^1_{NN}(\lambda)-\Phi^1_{ND}(\lambda)\Phi^1_{DN}(\lambda)}{\Delta_1(\lambda)}\Delta_2(\lambda)+
$$
$$
\Phi^1_{NN}(\lambda)\Phi^2_{DD}(\lambda)+\Phi^2_{NN}(\lambda)\Phi^1_{DD}(\lambda)+\Phi^1_{ND}(\lambda)\Phi^2_{DN}(\lambda)+\Phi^2_{ND}(\lambda)\Phi^1_{DN}(\lambda)+
$$
$$
\frac{\Phi^2_{DD}(\lambda)\Phi^2_{NN}(\lambda)-\Phi^2_{ND}(\lambda)\Phi^2_{DN}(\lambda)}{\Delta_2(\lambda)}\Delta_1(\lambda).
$$
Due to (\ref{5.11}) and (\ref{5.12}) we conclude that $\Delta_1(\lambda)\Delta_2(\lambda)D(\lambda)$ is a meromorphic  function with the set of zeros which coincides with the set of zeros of the entire function $\Phi_{NN}(\lambda)$, i.e. the characteristic function of the Neumann-Neumann problem on the whole graph. To prove this we notice that 
$$
|\Phi_{NN}(\lambda)|\mathop{=}_{|Im \lambda|\rightarrow \infty} C|Im\lambda|^{-p_1-p_2-2+q_1+q_2} e^{|Im \lambda| \ L}(1+o(1))
$$
where $L=L^{(1)}+L^{(2)}=\sum_{k=1}^{g_1}l_k^{(1)}+\sum_{k=1}^{g_2}l_k^{(2)}$, $p_j$ is the number of vertices in $G_j$, and the number of vertices in $G$ is $p_1+p_2-2$.
Due to (\ref{5.11}) and (\ref{5.13}) we obtain
$$
|\Delta_1(\lambda)\Delta_2(\lambda)D(\lambda)|\mathop{=}_{|Im \lambda|\rightarrow \infty} \tilde{C}|Im\lambda|^{-p_1-p_2-2+q_1+q_2} e^{|Im \lambda| \ L}(1+o(1)).
$$

We have proved the following theorem.
 
{\bf Theorem} {\it For a graph consisting of t parallel connected subgraphs
$$
\Phi_{NN}(\lambda)=\Delta_1(\lambda)\Delta_2(\lambda)D(\lambda)
$$
where $D(\lambda)$ is given by (\ref{5.8}).}

If we connect $m$ parallel subgraphs we obtain
\newpage
\begin{equation}
\label{5.13}
\Phi_{NN}=
\end{equation}
$$
\left|\begin{array}{cccccccccccc}
0&0&...&...&...&0&1&-1&0&...&...&0\\
0&0&...&...&...&0&1&0&-1&...&...&0\\
...&...&...&...&...&...&...&...&...&...&...&...\\
...&...&...&...&...&...&...&...&...&...&...&...\\
0&0&...&...&...&0&1&0&...&...&0&-1\\
1&1&...&...&...&1&0&0&...&...&...&0\\
\Phi^1_{DD}&-\Phi^2_{DD}&0&...&...&0&\Phi^1_{ND}&-\Phi^2_{ND}&0&...&...&0\\
\Phi^1_{DD}&0&-\Phi^3_{DDwo }&0&...&0&\Phi^1_{ND}&0&\Phi^2_{ND}&0&...&0\\
...&...&...&...&...&...&...&...&...&...&...&...\\
...&...&...&...&...&...&...&...&...&...&...&...\\
\Phi^1_{DD}&0&...&...&...&\Phi^m_{DD}&\Phi^1_{ND}&0&...&...&...&\Phi^m_{ND}\\
\Phi^1_{DN}&\Phi^2_{DN}&...&...&...&\Phi^m_{DN}&\Phi^1_{NN}&\Phi^2_{NN}&...&...&...&\Phi^m_{NN}
\end{array}\right|\mathop{\prod}\limits_{j=1}^{m}{\Delta_j}.
$$

Formulae (\ref{3.1})-(\ref{3.5}) and (\ref{5.i})-(\ref{5.iii}), (\ref{5.13}) remain true for problems generated by finite dimensional  analogue of Sturm-Liouville equation, so-called Stieltjes string equation. 

\vspace{3mm}


\end{document}